# Chapter Number

# Ferroelectricity in Silver Perovskite Oxides

Desheng Fu[1] and Mitsuru Itoh[2]
*[1]Shizuoka University,*
*[2]Tokyo Institute of Technology*
*Japan*

## 1. Introduction

Many ferroelectric oxides possess the $ABO_3$ perovskite structure (Mitchell, 2002), in which the A-site cations are typically larger than the B-site cations and similar in size to the oxygen anion. Figure 1 shows a schematic drawing for this structure, where the A cations are surrounded by 12-anions in the cubo-octahedral coordination and the B cations are surrounded by 6-anions in the octahedral coordination. An ideal perovskite exhibits a cubic space group *Pm3m*. This structure is centrosymmetric and cannot allow the occurrence of ferroelectricity that is the presence of a switchable spontaneous electric polarization arising from the off-center atomic displacement in the crystal (Jaffe et al., 1971; Lines & Glass, 1977). The instability of ferroelectricity in the perovskite oxides is generally discussed with the Goldschmidt tolerance factor ($t$) (Goldschmidt, 1926 and Fig. 2),

$$t = (r_A + r_O) / \sqrt{2}(r_B + r_O),$$

where $r_O$, $r_A$, and $r_B$ are the ionic radii of the O, A, and B ions. For a critical value $t=1$, the cubic paraelectric phase is stable. This unique case can be found in $SrTiO_3$, which has an ideal cubic perovskite structure at room temperature and doesn't show ferroelectricity down to the absolute 0 K (Müller & Burkard, 1979). However, ferroelectricity can be induced by the substitution of $O^{18}$ in this quantum paraelectric system at $T<T_c\sim23$ K (Itoh et al., 1999). When $t>1$, since the B-site ion is too small for its site, it can shift off-centeringly, leading to the occurrence of displacive-type ferroelectricity in the crystal. Examples of such materials are $BaTiO_3$ and $KNbO_3$ (Shiozaki et al., 2001). On the other hand, for $t<1$, the perovskite oxides are in general not ferroelectrics because of different tilts and rotations of $BO_6$ octahedra, which preserve the inversion symmetry. But exceptions may be found in the Bi-based materials, in which large A-site displacement is observed. This large A-site displacement is essentially attributed to the strong hybridization of Bi with oxygen (Baettig et al., 2005). Similar cases are observed in Pb-based materials, which commonly have large Pb displacement in the A-site (Egami et al., 1998) and strong covalent nature due to the unique stereochemistry of Pb (Cohen, 1992; Kuroiwa et al., 2001).

Although $BaTiO_3$- and $PbTiO_3$-based ceramics materials have been widely used in electronic industry (Uchino, 1997; Scott, 2000), there remain some importance issues to be solved. One of such challenges is to seek novel compounds to replace the Pb-based materials, which have a large Pb-content and raises concerns about the environmental pollution (Saito et al. ,2004; Rodel et al.,2009).

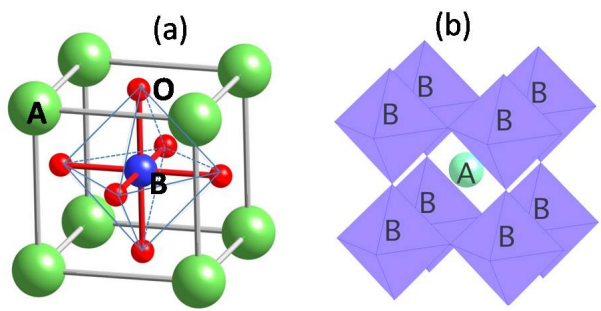

Fig. 1. The structure of an ABO₃ perovskite with the origin centered at (a) the B-site ion and (b) the A-site ion.

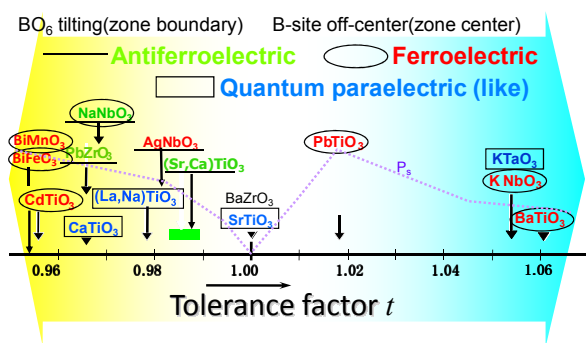

Fig. 2. Tolerance factor of typical dielectric oxides.

The discovery of extremely large polarization (52 µC/cm²) under high electric field in the AgNbO₃ ceramics (Fu et al., 2007) indicates that Ag may be a key element in the designs of lead-free ferroelectric perovskite oxides (Fu et al., 2011). With the advance of first-principles calculations (Cohen, 1992) and modern techniques of synchrotron radiation (Kuroiwa et al., 2001), we now know that the chemical bonding in the perovskite oxides is not purely ionic as we have though, but also possesses covalent character that plays a crucial role in the occurrence of ferroelectricity in the perovskite oxides (Cohen, 1992; Kuroiwa et al., 2001 ). It is now accepted that it is the strong covalency of Pb with O that allows its large off-center in the A-site. Although Ag doesn't have lone-pair electrons like Pb, theoretical investigations suggest that there is hybridization between Ag and O in AgNbO₃(Kato et al., 2002; Grinberg & Rappe, 2003,2004), resulting in a large off-center of Ag in the A-site of perovskite AgNbO₃ (Grinberg & Rappe, 2003,2004). This prediction is supported by the results from X-ray photoelectron spectroscopy, which suggest some covalent character of the chemical bonds between Ag and O as well as the bonds between Nb and O (Kruczek et al., 2006). Moreover, bond-length analysis also supports such a theoretical prediction. Some of the bond-lengths (~2.43 Å) in the structure (Sciau et al., 2004; Yashima et al. 2011) are significantly less than the sum of Ag⁺ (1.28 Å) and O²⁻ (1.40 Å) ionic radii (Shannon, 1967). All these facts make us believe that AgNbO₃ may be used as a base compound to develop novel ferroelectric materials. Along such a direction, some interesting results have been obtained. It was found that ferroelectricity can be induced through the chemical modification of the $AgNbO_3$ structure by substitution of Ag with Li (Fu et al., 2008, 2011), Na(Arioka, 2009; Arioka et al., 2010), and K (Fu et al., 2009a). Large spontaneous polarization and high temperature of para-ferroelectric phase transition were observed in these solid solutions. In the following sections, we review the synthesis, structure, and dielectric, piezoelectric and ferroelectric properties of these solid solutions together with another silver perovskite $AgTaO_3$ (Soon et al.,2009, 2010), whose solid solutions with $AgNbO_3$ are promising for the applications in microwaves devices due to high dielectric constant and low loss (Volkov et al. 1995; Fortin et al., 1996; Petzelt et al., 1999; Valant et al., 2007a).

## 2. AgNbO₃

### 2.1 Synthesis

Both single crystal and ceramics of $AgNbO_3$ are available. Single crystal can be grown by a molten salt method using NaCl or $V_2O_5$ as a flux (Łukaszewski et al., 1980; Kania, 1989). Ceramics samples can be prepared through a solid state reaction between $Nb_2O_5$ and silver source (Francombe & Lewis, 1958; Reisman & Holtzberg, 1958). Among the silver sources of metallic silver, $Ag_2O$ and $Ag_2CO_3$, $Ag_2O$ is mostly proper to obtain single phase of $AgNbO_3$(Valant et al., 2007b). For silver source of $Ag_2O$, thermogravimetric analysis indicates that phase formation can be reach at a firing temperature range of 1073-1397 K (Fig. 3). The issue frequently encountered in the synthesis of $AgNbO_3$ is the decomposition of metallic silver, which can be easily justified from the color of the formed compounds. Pure powder is yellowish, and grey color of the powder generally indicates the presence of some metallic silver. It has been shown that the most important parameter that influences the phase formation is oxygen diffusion (Valant et al., 2007b). In our experiments to prepare the $AgNbO_3$ ceramics, we first calcined the mixture of $Ag_2O$ and $Nb_2O_5$ at 1253 K for 6 hours in $O_2$ atmosphere and then sintered the pellet samples for electric measurements at 1323 K for 6 hours in $O_2$ atmosphere (Fu et al., 2007). Insulation of these samples is very excellent, which allows us to apply extremely high electric field to the sample (Breakdown field >220 kV/cm. For comparison, $BaTiO_3$ ceramics has a value of ~50 kV/cm.)

### 2.2 Electric-field induced ferroelectric phase

Previous measurements on *D-E* hysteresis loop by Kania *et al.* (Kania et al., 1984) indicate that there is small spontaneous polarization $P_s$ in the ceramics sample of $AgNbO_3$. $P_s$ was estimated to be ~0.04 μC/cm² for an electric field with an amplitude of $E$=17 kV/cm and a frequency of 60 Hz. Our results obtained at weak field have confirmed Kania's reports (Fig. 4 and Fu et al., 2007). The presence of spontaneous polarization indicates that $AgNbO_3$ must be ferroelectric at room temperature. The good insulation of our samples allows us to reveal a novel ferroelectric state at higher electric field. As shown in Fig.4, double hysteresis loop is distinguishable under the application of $E$~120 kV/cm, indicating the appearance of new ferroelectric phase. When $E$>150 kV/cm, phase transformation is nearly completed and very large polarization was observed. At an electric field of $E$=220 kV/cm, we obtained a value of 52 μC/cm² for the ceramics sample. Associating with such structural change, there is very large electromechanical coupling in the crystal. The induced strain was estimated to be 0.16% for the ceramics sample (Fig.5). The *D-E* loop results unambiguously indicate that the atomic displacements are ordered in a *ferri*-electric way rather than an ***anti-ferro***electric way in the crystal at room temperature.

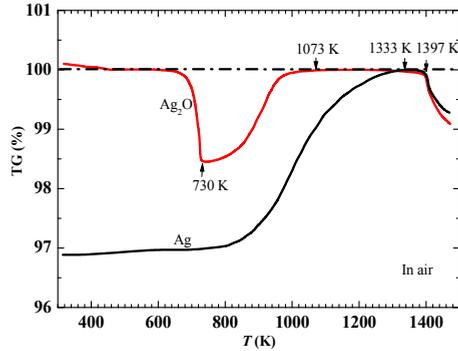

1
2    Fig. 3. Thermogravimetric curves of the AgNbO$_3$ formation in air using Ag$_2$O or metallic Ag
3    powder as the silver source (Valant et al., 2007b). The curves are normalized to a weight of
4    single-phase AgNbO$_3$. Temperatures of phase formation completed and decomposition are
5    also indicated. For case of Ag$_2$O, decomposition of Ag$_2$O into Ag and oxygen occurs at
6    temperature of ~730 K.
7

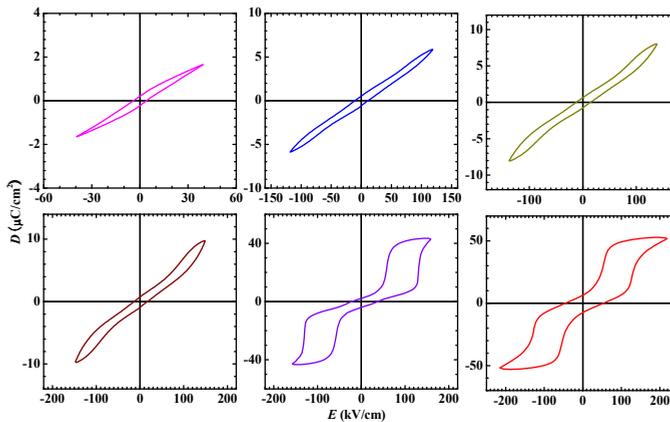

8
9    Fig. 4. D-E hysteresis loops for poly-crystalline AgNbO$_3$ at room temperature.
10

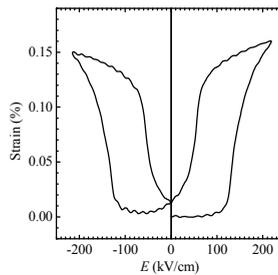

11
12   Fig. 5. Strain vs electric field for poly-crystalline AgNbO$_3$ at room temperature.

### 2.3 Room-temperature structure

There are many works attempting to determine the room-temperature structure of $AgNbO_3$ (Sciau et al., 2004; Francombe & Lewis, 1958; Verwerft et al., 1989; Fabry et al., 2000; Levin et al., 2009). However, none of these previous works can provide a *non*-centrosymmetric structure to reasonably explain the observed spontaneous polarization. Very recently, this longstanding issue has been addressed by R. Sano *et al.* (Sano et al., 2010). The space group of $AgNbO_3$ has been unambiguously determined to be $Pmc2_1$ (No. 26) by the convergent-beam electron diffraction (CBED) technique, which is *non*-centrosymmetric and allows the appearance of ferroelectricity in the crystal (Fig.6).

| $Pmc2_1$ ($T$=298K) | | | | |
|---|---|---|---|---|
| Site | $x$ | $y$ | $z$ | $U$ (Å²) |
| Ag1 $4c$ | 0.7499(3) | 0.7468(3) | 0.2601(5) | 0.0114(2) |
| Ag2 | 1/2 | 0.7466(6) | 0.2379(5) | 0.0114(2) |
| Ag3 | 0 | 0.7424(4) | 0.2759(6) | 0.0114(2) |
| Nb1 | 0.6252(2) | 0.7525(5) | 0.7332(2) | 0.00389(18) |
| Nb2 | 0.1253(2) | 0.24159 | 0.27981 | 0.00389(18) |
| O1 $4c$ | 0.7521(9) | 0.7035(12) | 0.7832(24) | 0.0057(5) |
| O2 $2b$ | 1/2 | 0.804(3) | 0.796(3) | 0.0057(5) |
| O3 $4c$ | 0.6057(7) | 0.5191(18) | 0.4943(18) | 0.0057(5) |
| O4 $4c$ | 0.6423(7) | 0.0164(18) | 0.539(2) | 0.0057(5) |
| O5 $2a$ | 0 | 0.191(3) | 0.256(3) | 0.0057(5) |
| O6 $4c$ | 0.1339(9) | 0.0410(17) | 0.980(2) | 0.0057(5) |
| O7 $4c$ | 0.1154(8) | 0.4573(17) | 0.5514(19) | 0.0057(5) |

Table 1. Structural parameters of $AgNbO_3$ at $T$=298K (Yashima et al., 2011). Number of formula units of $AgNbO_3$ in a unit cell: $Z$=8. Unit-cell parameters: $a$ = 15.64773(3) Å, $b$ = 5.55199(1) Å, $c$ = 5.60908(1) Å, $\alpha = \beta = \gamma = 90$ deg., Unit-cell volume: $V$ = 487.2940(17) Å³. $U$ (Å²)=Isotropic atomic displacement parameter.

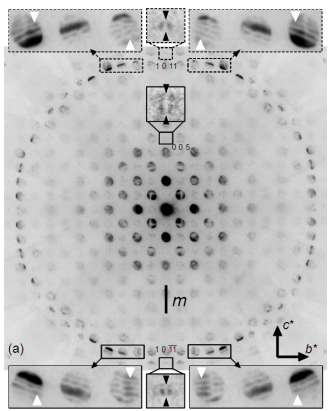

Fig. 6. Convergent-beam electron diffraction (CBED) pattern of $AgNbO_3$ taken at the [100] incidence. In contrast to a mirror symmetry perpendicular to the $b^*$-axis, breaking of mirror symmetry perpendicular to the $c^*$-axis is seen, indicating that spontaneous polarization is along the $c$-direction (Taken by R. Sano & K. Tsuda (Sano et al., 2010)).

On the basis of this space group, M. Yashima (Yashima et al., 2011) exactly determined the atom positions (Table 1) in the structure using the neutron and synchrotron powder diffraction techniques. The atomic displacements are schematically shown in Fig.7. In contrast to the reported centrosymmetric *Pbcm* (Fabry et al. 2000; Sciau et al. 2004; Levin et al. 2009), in which the Ag and Nb atoms exhibit antiparallel displacements along the *b*-axis, the *Pmc*2$_1$ structure shows a ***ferri*-electric** ordering of Ag and Nb displacements (Yashima et al., 2011) along the *c*-axis of *Pmc*2$_1$ orthorhombic structure. This polar structure provides a reasonable interpretation for the observed polarization in AgNbO$_3$.

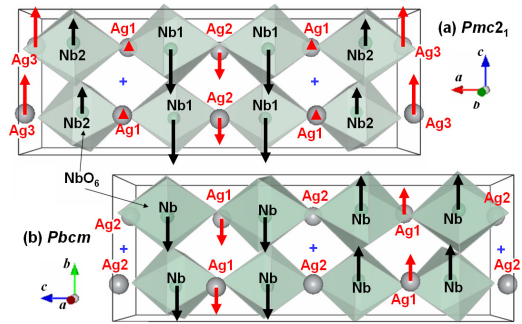

Fig. 7. (a) *Ferri*electric crystal structure of AgNbO$_3$ (*Pmc*2$_1$) at room temperature. The atomic displacements along the *c*-axis lead to the spontaneous polarization in the crystal. (b) For comparison, the patterns for the previously reported *Pbcm* (Sciau et al., 2004) are also given. A cross (+) stands for the center of symmetry in the *Pbcm* structure. (by M. Yashima (Yashima et al., 2011) ).

## 2.4 Dielectric behaviours and phase transitions

Initial works on the phase transitions of AgNbO$_3$ and their influence on the dielectric behaviors were reported by Francombe and Lewis (Francombe & Lewis, 1958) in the late 1950s, which trigger latter intensive interests in this system (Łukaszewski et al., 1983; Kania, 1983, 1998; Kania et al., 1984, 1986; Pisarski & Dmytrow, 1987; Paweczyk, 1987; Hafid et al., 1992; Petzelt et al., 1999; Ratuszna et al., 2003; Sciau et al., 2004). The phase transitions of AgNbO$_3$ were associated with two mechanisms of *displacive* phase transition: tilting of oxygen octahedra and displacements of particular ions (Sciau et al., 2004). Due to these two mechanisms, a series of structural phase transitions are observed in AgNbO$_3$. The results on dielectric behaviors together with the reported phase transitions are summarized in Fig.8. Briefly speaking, the structures of the room-temperature (Yashima et al., 2011) and the high temperature phases (*T*> *T*$^{O1-O2}$=634 K ) (Sciau et al., 2004) are exactly determined, in contrast, those of low-temperature (*T*<room temperature) and intermediate phases ( *T*$_C$$^{FE}$=345 K <*T*< *T*$^{O1-O2}$) remain to be clarified. In the dielectric curve, we can see a shoulder around 40 K. It is unknown whether this anomaly is related to a phase transition or not. It should be noticed that Shigemi *et al.* predicted a ground state of *R3c* rhombohedra phase similar to that of NaNbO$_3$ for AgNbO$_3$ (Shigemi & Wada, 2008). Upon heating, there is an anomaly at *T*$_C$$^{FE}$=345 K, above which spontaneous polarization was reported to disappear (Fig.8 (c) and Kania et al., 1984). At the same temperature, anomaly of lattice distortion was observed (Fig.8 (c) and Levin et al., 2009). The dielectric anomaly at *T*$_C$$^{FE}$=345 K was attributed to be a ferroelectric phase transition. Upon further heating, there is a small peak at *T*=453 K, which is not so visible. However, it can be easily ascertained in the differential curve or in the cooling curve. This anomaly is nearly unnoticed in the literatures (Łukaszewski et al., 1983; Kania, 1983, 1998; Kania et al., 1986; Pisarski & Dmytrow, 1987; Paweczyk, 1987; Hafid et al., 1992; Ratuszna et al., 2003). The detailed examination of the temperature dependence of the 220$_o$ *d*-spacing (reflection was indexed with orthorhombic structure) determined by Levin *et al.* (Levin et al., 2009 and Fig.8(c)) reveals anomaly that can be associated with this dielectric peak. These facts suggest that a phase transition possibly occurs at this temperature. Around 540 K, there is a broad and frequency-dependent peak of dielectric constant, which is also associated with an anomaly of 220$_o$ *d*-spacing. However, current structural investigations using x-ray and neutron diffraction do not find any symmetric changes associated with the dielectric anomalies at 456 K and 540 K, and the structure within this intermediate temperature range was assumed to be orthorhombic *Pbcm* (Sciau et al., 2004). At $T=T_C{}^{AFE}=631$ K, there is a sharp jump of dielectric constant due to an antiferroelectric phase transition (Pisarski et al., 1987; Sciau et al., 2004). The atomic displacement patterns in the antiferroelectric phase (*Pbcm*) are shown in Fig.7 (b) using the structural parameters determined by Ph Sciau *et al.* (Sciau et al., 2004). For $T>T_C{}^{AFE}$, there are still three phase transitions that are essentially derived by the tilting of oxygen octahedral and have only slight influence on the dielectric constant. For conveniences, the tilting of octahedral (Sciau et al., 2004) described in Glazer's notation is given in Fig.8 and the structure parameters (Sciau et al., 2004) are relisted in Table 2.

| | | AgNbO$_3$ | | | |
|---|---|---|---|---|---|
| Atom | | 573 K | 645 K | 733 K | 903 K |
| Nb | $x$ | 0.2460(15) | 1/4 | 0 | 0 |
| | $y$ | 0.2422(10) | 1/4 | 0 | 0 |
| | $z$ | 0.1256(6) | 0 | 0 | 0 |
| | $U$ (Å$^2$) | 1.12(6) | 1.31(4) | 1.38(4) | 1.63(2) |
| Ag(1) | $x$ | -0.2507(35) | 0 | 0 | 1/2 |
| | $y$ | 1/4 | -0.001(2) | 1/2 | 1/2 |
| | $z$ | 0 | 1/4 | 1/2 | 1/2 |
| | $U$ (Å$^2$) | 2.78(7) | 1.98(6) | 2.80(3) | 3.47(4) |
| Ag(2) | $x$ | -0.2556(26) | 0 | | |
| | $y$ | 0.2428(18) | 0.494(3) | | |
| | $z$ | 1/4 | 1/4 | | |
| | $U$ (Å$^2$) | 1.73(6) | 3.54(8) | | |
| O(1) | $x$ | 0.3022 | 0.2827(7) | 0 | 0 |
| | $y$ | 1/4 | 1/4 | 0 | 0 |
| | $z$ | 0 | 0 | 1/2 | 1/2 |
| | $U$ (Å$^2$) | 1.31(6) | 2.27(5) | 3.14(5) | 3.60(4) |
| O(2) | $x$ | -0.0292(14) | 0 | 0.2782(3) | |
| | $y$ | 0.0303(14) | 0.2259(8) | 0.2218 | |
| | $z$ | 0.1140(6) | 0.0205(8) | 0.0358(4) | |
| | $U$ (Å$^2$) | 1.71(7) | 1.94(4) | 2.08(4) | |
| O(3) | $x$ | 0.5262(13) | 0.2710(7) | | |
| | $y$ | 0.4778(14) | 0.2456(11) | | |
| | $z$ | 0.1375(5) | 1/4 | | |
| | $U$ (Å$^2$) | 1.17(6) | 2.56(6) | | |
| O(4) | $x$ | 0.2155(23) | | | |
| | $y$ | 0.2622(24) | | | |
| | $z$ | 1/4 | | | |
| | $U$ (Å$^2$) | 1.67(7) | | | |
| cell | $a$ (Å) | 5.5579(5) | 7.883(1) | 5.5815(3) | 3.9598(3) |
| | $b$ (Å) | 5.5917(6) | 7.890(1) | | |
| | $c$ (Å) | 15.771(2) | 7.906(1) | 3.9595(3) | |

Table 2. Structural parameters for high temperature phases at 573 K(*Pbcm*), 645 K(*Cmcm*), 733 K (*P4/mbm*), and 903 K(*Pm3m*) (Sciau et al., 2004).

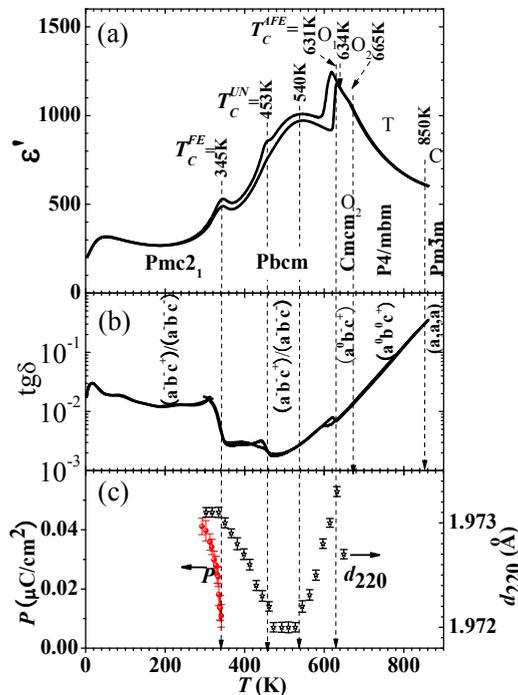

Fig. 8. Temperature dependence of (a) dielectric constant, (b) dielectric loss, and (c) polarization (Kania et al., 1984) and $220_O$ $d$-spacing of the lattice (Levin et al., 2009).

## 3. (Ag$_{1-x}$Li$_x$)NbO$_3$ solid solution

Li can be incorporated into the Ag-site of AgNbO$_3$. However, due to the large difference of ionic radius of Li$^+$(0.92Å), and Ag$^+$( 1.28Å)(Shannon, 1976), the solid solution is very limited. Nalbandyan *et al.*(Nalbandyan et al., 1980), systematically studied the stable and metastable phase equilibrias and showed that solid solution limit is narrow (*x*~0.02) for the stable phase, but is relatively wide (*x*~0.12) for the metastable phase(Sakabe et al., 2001;

Takeda et al., 2003; Fu et al., 2008, 2011). With a small substitution of Li for Ag ($x>x_c$=0.05~0.06), a ferroelectric rhombohedra phase is evolved in the solid solution (Nalbandyan et al., 1980; Sakabe et al., 2001; Takeda et al., 2003;Fu et al., 2008, 2011). In this solid solution, the strong off-center of small Li (Bilc & Singh, 2006) plays an important role in triggering the ferroelectric state with large spontaneous polarization (Fu et al., 2008, 2011).

### 3.1 Synthesis

Single crystals of $(Ag_{1-x}Li_x)NbO_3$ can be obtained by the melt growth process (Fu et al., 2008). Stoichiometric compositions of $Ag_2O$, $Li_2CO_3$, and $Nb_2O_5$ were mixed and calcined at 1253 K for 6 h in an oxygen atmosphere. The calcined powder was milled, put into an alumina container, and melted at 1423 K for 4 h in an oxygen atmosphere. The melt was cooled to 1323 K to form the crystal at a rate of 4 K/h, followed by furnace cooling down to room temperature. Using this process, single crystals with size of 1–3 cm can be obtained for the $(100)_p$ (Hereafter, subscript $p$ indicates pseudocubic structure) growth face. Due to the volatility of lithium at high temperature, the exact chemical composition of the crystal is generally deviated from the starting composition and is required to be determined with methods like inductively coupled plasma spectrometry.

Ceramics samples can be prepared by a solid state reaction approach. Mixtures of $Ag_2O$, $Nb_2O_5$, and $Li_2CO_3$ were calcined at 1253 K for 6 h in $O_2$ atmosphere, followed by removal of the powder from the furnace to allow a rapid cooling to prevent phase separation. The calcined powder was milled again and pressed to form pellets that were sintered at 1323 K for 6 h in $O_2$ atmosphere, followed by a rapid cooling.

### 3.2 Structure

The structural refinements using the powder X-ray diffraction data suggest that $(Ag_{1-x}Li_x)NbO_3$ solid solution with $x>x_c$ has the space group of $R3c$ (Fu et al., 2011). Table 3 lists the structural parameters of this model for composition $x$=0.1. Figure 9 shows a schematic drawing for this structure. In this rhombohedal $R3c$ phase, the spontaneous polarization is essentially due to the atomic displacements of the Ag/Li, Nb, and O atoms along the pseudocubic [111] direction.

| $Ag_{0.9}Li_{0.1}NbO_3$ (R3c, No.161, T=room temperature) | | | | | |
|---|---|---|---|---|---|
| $a$(Å) | 5.520 | | α | 90 | |
| $b$(Å) | 5.520 | | β | 90 | |
| $c$(Å) | 13.79 | | γ | 120 | |
| $V$(Å$^3$) | 364.0 | | | | |
| Atom | Site | $x$ | $y$ | $z$ | $U$ (Å$^2$) |
| Ag/Li | 6a | 0 | 0 | 0.2545(8) | 0.5 |
| Nb | 6a | 0 | 0 | 0.0097(8) | 0.5 |
| O | 18b | 0.5533 | 1 | 0.2599(9) | 0.5 |

Table 3. Structural parameters for rhombohedra structure of $(Ag,Li)NbO_3$ solid solution.

### 3.3 Ferroelectric and piezoelectric properties

Evolution of the polarization state in $Ag_{1-x}Li_xNbO_3$ solid solutions is shown in Fig.10. Basically, when $x<x_c$, the solid solutions have the *ferri*electric state of pure $AgNbO_3$ with a small spontaneous polarization at zero electric field. In contrast, when $x>x_c$, a normal ferroelectric state with large value of remanent polarization ($P_r$) is observed. All ceramics samples show $P_r$ value comparable to $P_S$ of $BaTiO_3$ single crystal (26 $\mu C/cm^2$) (Shiozaki et al., 2001). Moreover, the polarization in $Ag_{1-x}Li_xNbO_3$ solid solution is very stable after switching. Large $P_r$ value and ideal bistable polarization state of $Ag_{1-x}Li_xNbO_3$ ceramics may be interesting for non-volatile ferroelectric memory applications. Measurements on single crystal samples (Fig.11) indicate that saturation polarization along the <111>$_p$ rhombohedra direction ($P_s^{<111>}$ ~40 $\mu C/cm^2$) is greatly larger than that along the <001>$_p$ tetragonal direction ($P_s^{<001>}$ ~24 $\mu C/cm^2$) and the ratio between them is $\sqrt{3}$ (Fu et al., 2008), which is in good agreement with results of structural refinements. This suggests that the polar axis is the <111>$_p$ direction of pseudo-cubic structure.

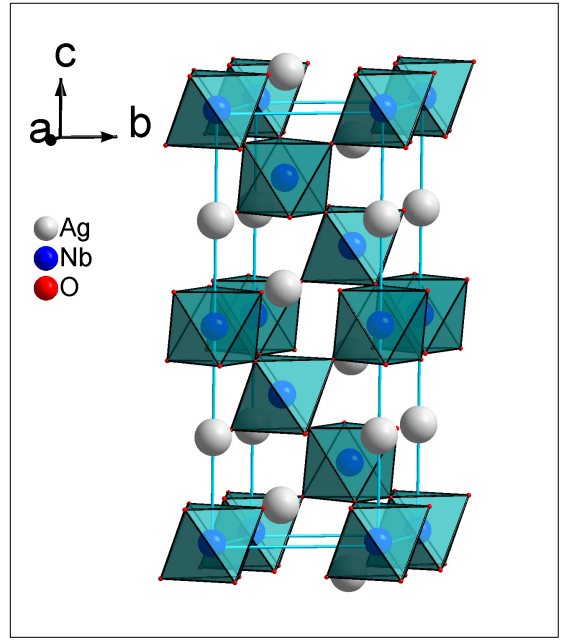

Fig. 9. Schematic structure of $Ag_{1-x}Li_xNbO_3$ ($x$ = 0.1) with symmetry $R$3c (No.161).

The strain-$E$ loops indicate that there are good electromechanical coupling effects in $Ag_{1-x}Li_xNbO_3$ crystals. Although the spontaneous polarization is along the <111>$_p$ axis, the <001>$_p$-cut crystal shows larger strain and less hysteresis than the <111>$_p$-cut one (Fig.11 (b) and (c)). These phenomena are very similar to those reported for the relaxor-ferroelectric crystals (Wada et al., 1998). The most significant result exhibited from $Ag_{1-x}Li_xNbO_3$ single crystal is its excellent $g_{33}$ value that determines the voltage output of the piezoelectric device under the application of an external stress (Fu et al., 2008). The $g_{33}$ value together the $d_{33}$ value and dielectric constants for the <001>$_p$-cut single crystals are shown in Fig. 12. The high $g_{33}$ value is a direct result from the large $d_{33}$ constant and the low dielectric constant of the single crystal.

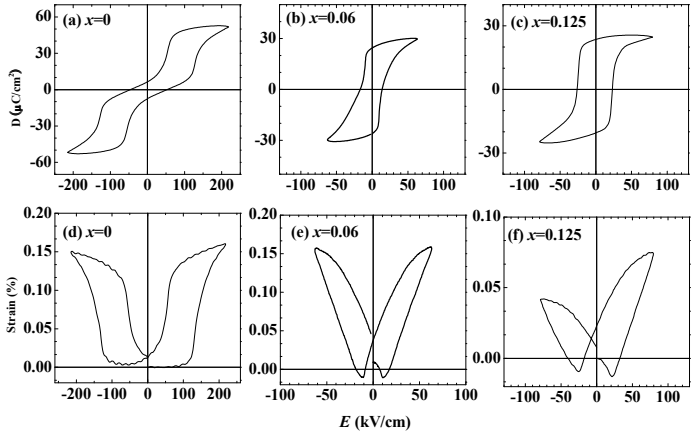

Fig. 10. Typical *D-E* and strain-*E* loops for the Ag$_{1-x}$Li$_x$NbO$_3$ ceramics samples.

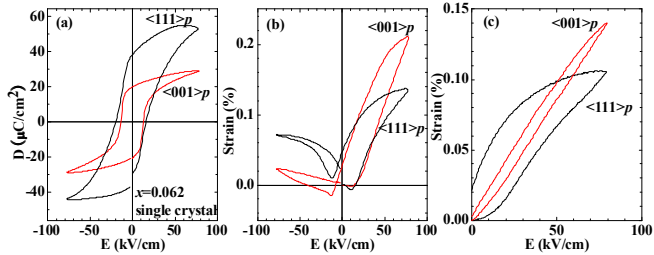

Fig. 11. (a) *D-E* loops, (b) strain vs *E* for bipolar electric field, and (c) strain vs *E* for unipolar field for Ag$_{1-x}$Li$_x$NbO$_3$ single crystal with *x*=0.062.

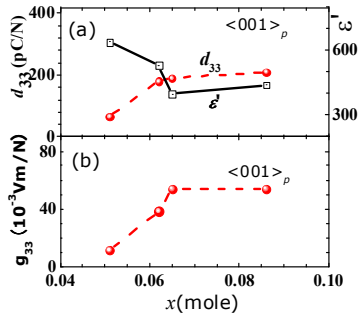

Fig. 12. Composition dependence of dielectric constant $\varepsilon$, $d_{33}$ and $g_{33}$ for the <001>$_p$-cut Ag$_{1-x}$Li$_x$NbO$_3$ single crystals.

### 3.4 Dielectric behaviours and proposed phase diagram

Figure 13 shows the dielectric behaviours of the ferroelectric Ag$_{1-x}$Li$_x$NbO$_3$ solid solutions. For comparison, the temperature dependence of dielectric constant of AgNbO$_3$ is also shown. It can be seen that solid solution with $x > x_c$ shows different temperature evolutions of the dielectric constant as compared with AgNbO$_3$. In sharp contrast to the complex successive phase transitions in AgNbO$_3$, ferroelectric Ag$_{1-x}$Li$_x$NbO$_3$ solid solutions ($x > x_c$) show only two phase transitions in the temperature range of 0-820 K. Polarization measurements suggest that the high temperature phase ($T > T_C^{FE}$) is nonpolar, thus it seems that the higher temperature phase transition is not related to a ferroelectric phase transition. On the basis of the dielectric measurements, the phase diagram of $Ag_{1-x}Li_xNbO_3$ solution is summarized in Fig.14, where *T*, *O*, *R* and *M* represent tetragonal, orthorhombic, rhombohedra, and monoclinic symmetries, respectively. At room temperature, structure transformation from *O* to *R* phase at $x_c$ dramatically changes the polar nature of $Ag_{1-x}Li_xNbO_3$. It is a *ferri*electric with small spontaneous polarization in the *O* phase, but becomes a ferroelectric with large polarization in the *R* phase.

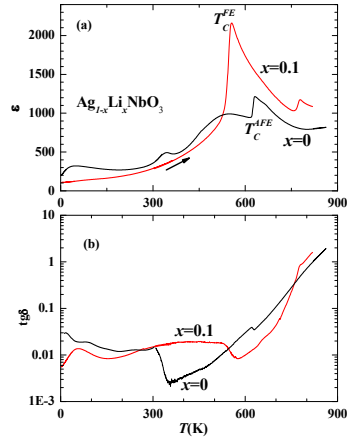

Fig. 13. Typical dielectric behaviours of Ag$_{1-x}$Li$_x$NbO$_3$ ceramics in comparison with that of pure AgNbO$_3$.

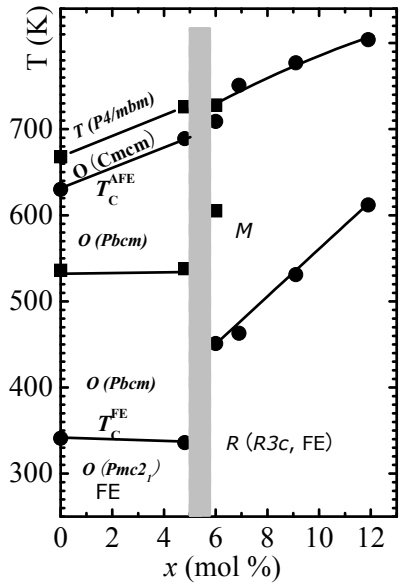

Fig. 14. Phase diagram proposed for Ag$_{1-x}$ Li$_x$NbO$_3$ solid solution. The gray area indicates the phase boundary (Fu et al., 2011).

## 4. (Ag$_{1-x}$Na$_x$)NbO$_3$ solid solution

The ionic radius of $Na^+$ (1.18 Å) is comparable to that of $Ag^+$ (1.28 Å) (Shannon, 1976), allowing to prepare the $Ag_{1-x}Na_xNbO_3$ solid solution within the whole range of composition (*x*=0-1) (Kania & Kwapulinski, 1999). Kania *et al*. previously performed investigation on the dielectric behaviors and the differential thermal analysis for the $Ag_{1-x}Na_xNbO_3$ solid solutions, and stated that the solid solution evolves from disordered antiferroelectric $AgNbO_3$ to normal antiferroelectric $NaNbO_3$. As described in section §2.3, we now know that $AgNbO_3$ is not *antiferroelectric* but rather is *ferrielectric* at room temperature (Yashima et al., 2011). Moreover, recent reexamination on the polarization behaviors of stoichiometric and non-stoichiometric $NaNbO_3$ polycrystallines indicates that the reported clamping hysteresis loop of $NaNbO_3$ can be interpreted by pining effects while stoichiometric $NaNbO_3$ is intrinsically *ferroelectric* (Arioka et al., 2010). Therefore, reexamination on this solid solution is necessary. Evolution of the polarization with composition clearly indicates that the solid solution evolves from *ferrielectric* $AgNbO_3$ to *ferroelectric* $NaNbO_3$.

### 4.1 Synthesis

$Ag_{1-x}Na_xNbO_3$ solid solution was prepared by a solid state reaction approach. Mixtures of $Ag_2O$(99%), $Nb_2O_5$ (99.99%), and $Na_2CO_3$ (99.99%) were calcined at 1173 K for 4 h in $O_2$ atmosphere. The calcined powder was ground, pressed into pellet with a diameter of 10 mm at thickness of 2 mm, and sintered with the conditions listed in Table 4.

| Composition | Temperature | Time | atmosphere |
|---|---|---|---|
| *x*=0,0.1,0.2 | 1273K | 5h | $O_2$ |
| *x*=0.4, 0.5, 0.6 | 1323K | 5h | $O_2$ |
| *x*= 0.8,0.9,1 | 1373K | 5h | $O_2$ |

Table 4. Sintering conditions for $Ag_{1-x}Na_xNbO_3$.

### 4.2 Polarization

Figure 15 shows the change in polarization with composition in the $Ag_{1-x}Na_xNbO_3$ solid solutions. For a wide range of composition *x*<0.8, the solid solution possesses the characteristic polarization behaviours of pure $AgNbO_3$: small spontaneous polarization at *E*=0 but large polarization at *E*> a critical field. This fact suggests that the solid solution is *ferri*electric within this composition range. This is also supported by the temperature dependence of dielectric constant (Fig.16). On the other hand, for the Na-rich composition, particularly, $x>0.8$, we observed large remanent polarization with value close to the saturation polarization at high field. This result indicates that a normal ferroelectric phase is stable in these compositions. Therefore, polarization measurements show that the $Ag_{1-x}Na_xNbO_3$ solid solution evolves from *ferrielectric* $AgNbO_3$ to *ferroelectric* $NaNbO_3$.

### 4.3 Dielectric properties

The dielectric properties of the $Ag_{1-x}Na_xNbO_3$ solid solutions are summarized in Fig.16. The change in dielectric behaviours with composition is very similar to that observed in polarization measurements (Fig.15). For $x \leq 0.8$, the solid solution shows successive phase transitions similar to pure $AgNbO_3$. In contrast, it has the characteristic phase transition of pure $NaNbO_3$ for $x>0.8$. The composition dependence of transition temperature derived from the dielectric measurements is shown in Fig.17. Two noticed features may be seen: (a) the thermal hysteresis is extremely large for antiferroelectric phase transition and reaches a value greater than 100 K at $x$~0.5. Such large thermal hysteresis is rarely observed in normal polar phase transition. (b) It seems that there is a phase boundary within $x$=0.8-0.9, around which structural transformation between *ferri*- and *ferro*-electric phases occurs.

## 5. $(Ag_{1-x}K_x)NbO_3$ solid solution

$(Ag_{1-x}K_x)NbO_3$ solid solutions are available only for very limited composition (Weirauch & Tennery, 1967; Łukaszewski, 1983; Kania, 2001). Weirauch *et al*. reported that solid solution of $AgNbO_3$ in $KNbO_3$ was limited to slightly less than 6 mole % and solid solution of $KNbO_3$ in $AgNbO_3$ was limited to less than 0.5 mole % (Weirauch & Tennery, 1967). However, our process indicates that $KNbO_3$ and $AgNbO_3$ can be alloyed with each other within 20 mole % (Fu et al., 2009a). Apparently, the reported solid solution limit is dependent on the process. In our samples, we found that a ferroelectric phase with large spontaneous polarization can be induced by substitution of Ag with K for $x>x_{c1}$=0.07. This ferroelectric phase shows nearly composition-independent ferroelectric phase transition. On the other hand, the K-rich solid solution ($x>0.8$) possesses the ferroelectric phase transition sequence of pure $KNbO_3$ and the transition temperature is dependent on the composition.

### 5.1 Synthesis

$(Ag_{1-x}K_x)NbO_3$ solid solutions were prepared by a solid state reaction method. Mixture of $Ag_2O$, $Nb_2O_5$, and $K_2CO_3$ were calcined at 1173 K for 6 h in $O_2$ atmosphere with a slow heating rate of 1 K/min. The calcined powder was milled again, pressed in a 6-mm steel die with a pressure of 10 MPa to form the pellets, which were then preheated at 773 K for 2 h, followed by a sintering at temperature of 1253–1323 K (1323 K for $x$=0–0.1 and 1.00, 1273 K for $x$=0.15 and 0.90, and 1253 K for $x$=0.17 and 0.80, respectively) for 3 h in $O_2$ atmosphere. The atmosphere and heating rate are found to have significant influences on the phase stability of the solid solution.

### 5.2 Structural change with composition

Figure 18 shows the change in lattice parameters with composition in the $Ag_{1-x}K_xNbO_3$ solid solutions. When the amount of substitution is small, the solid solution possesses the orthorhombic structure of pure $AgNbO_3$. In the phase boundary $x_{c1}$=0.07, structural transformation occurs and the phase changes into a new orthorhombic structure. In this new ferroelectric phase, the lattice constants show linear increase with composition. Interestingly, the orthorhombic distortion angle $\beta$ is nearly independent with the composition. In the K-rich region ($x > x_{c3} = 0.8$), the solid solution has the orthorhombic structure of pure KNbO$_3$ at room-temperature, which is also ferroelectric. In contrast to nearly unchanged orthorhombic angle $\beta$ in the Ag-rich orthorhombic phase, $\beta$ shows monotonous decreases with $x$ in the K-rich phase.

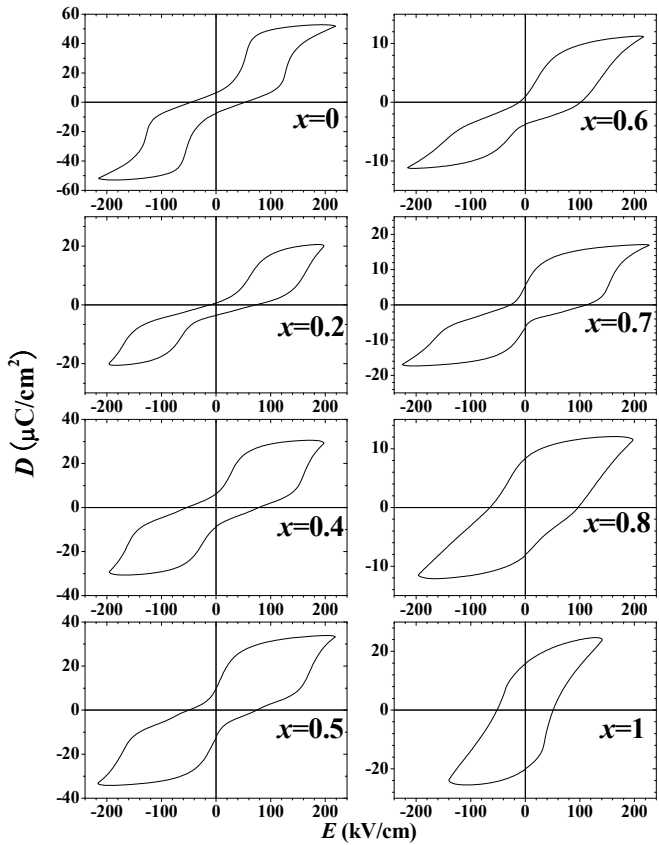

Fig. 15. Composition dependence of $D$-$E$ loops obtained at room temperature for Ag$_{1-x}$Na$_x$NbO$_3$ solid solutions

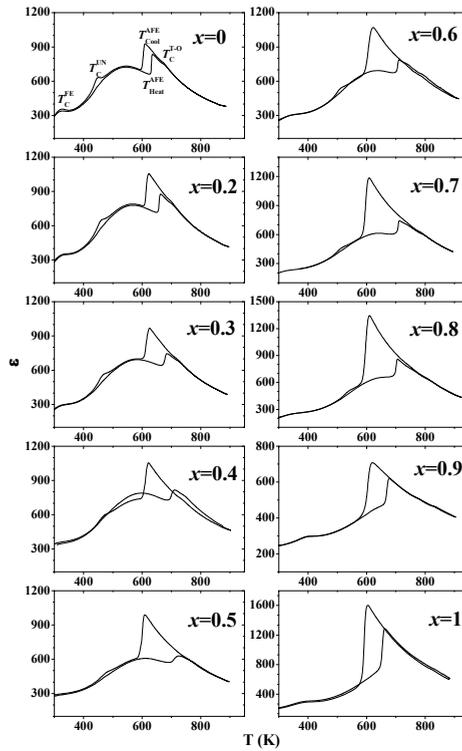

Fig. 16. Temperature dependence of dielectric constants for Ag$_{1-x}$Na$_x$NbO$_3$ solid solutions.

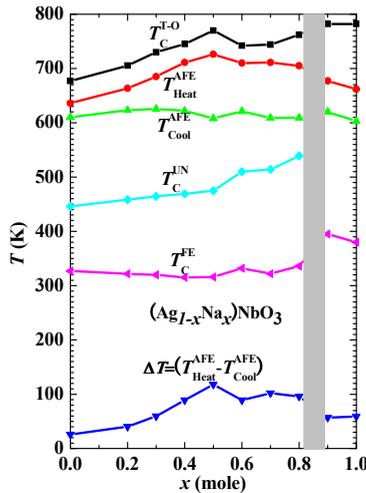

Fig. 17. Composition dependence of phase transition temperatures detected from dielectric measurements for Ag$_{1-x}$Na$_x$NbO$_3$ solid solutions.

## 5. 3 Ferroelectric and piezoelectric properties

Figure 19 shows typical results of polarization and strain behaviors observed at room temperature for the $Ag_{1-x}K_xNbO_3$ solid solutions. Similar to pure $AgNbO_3$, merely a small spontaneous polarization was observed in samples with $x<x_{c1}$. However, when $x>x_{c1}$, a normal $D$-$E$ loop with large value of remanent polarization $P_r$ was observed. A value of $P_r$=20.5 μC/cm² was obtained for a ceramics sample with $x$ =0.10, which is greatly larger than that observed for $BaTiO_3$ ceramics (Fu et al., 2010). These results show that Ag-rich orthorhombic phase ($x_{c1}<x<x_{c2}$ ) is really under a ferroelectric state with large polarization. For K-rich region $x>x_{c3}$, normal $D$-$E$ loops were also obtained. Associating with the evolution into the ferroelectric phase, butterfly strain curve were observed. The piezoelectric constants determined from the piezo-$d_{33}$ meter generally have value of 46–64 pC/N for these ferroelectric samples (Fu et al., 2009a).

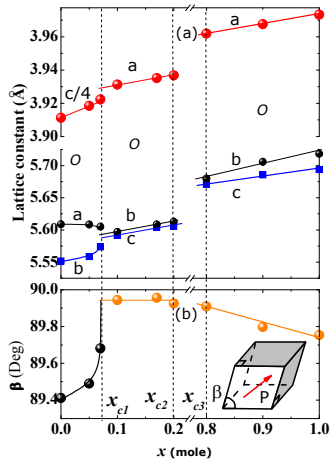

Fig. 18. Lattice parameters change with composition in $Ag_{1-x}K_xNbO_3$ solid solution. $\beta$ is the orthorhombic distortion angle. The inset shows the orthorhombic distortion due to the ferroelectric displacements along the <110>$_p$ direction of the pseudocubic structure.

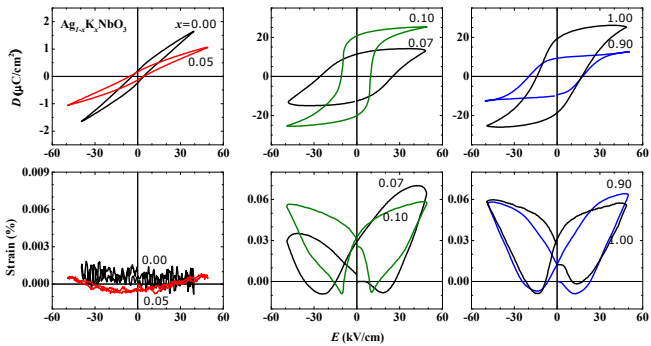

Fig. 19. Changes of polarization and strain under the application of electric field with composition at room temperature.

**5.4 Dielectric behaviours and proposed phase diagram**

Associating with the change in structure, dielectric behaviours of the $Ag_{1-x}K_xNbO_3$ solid solution also change with composition. As shown in Fig.20, the temperature dependence of dielectric constant can be sorted by three types: (1) $AgNbO_3$-type for $x<x_{c1}$=0.07, (2) $KNbO_3$-type for K-rich region $x>x_{c3}$=0.8, and (3) a new type for the intermediate composition $x_{c1}<x<x_{c2}$. In this intermediate composition, dielectric measurements indicate that there are two phase transitions within the temperature range of 0-750 K. One transition locates at $T_{c2}$~420 K with thermal hysteresis and shows small change in dielectric constant, which seems to be due to a ferro-to-ferro-electric phase transition. Another phase transition occurs at $T_{c1}$~525 K. Around this transition, the dielectric constant changes sharply and follows exactly the Curie-Weiss law. The Curie-Weiss constant was estimated to be 1.47 *10⁵ K for $x$=0.1 sample, which is a typical value for the displacive ferroelectric transition, suggesting that this is a displacive type ferroelectric transition. A phase diagram is proposed in Fig.21, in which PE, FE, and AFE represent the paraelectric, ferroelectric, and antiferroelectric phases, respectively. When carefully comparing the phase transition temperature with the orthorhombic angle β(Fig.18) due to the ferroelectric distortion (For $x_{c1}<x<x_{c2}$ and $x>x_{c3}$. In contrast, for $x<x_{c1}$, β change is basically due to the oxygen-octahedral tilting.) , one might find that there is a correlation between β and the temperature of the ferroelectric phase transition.

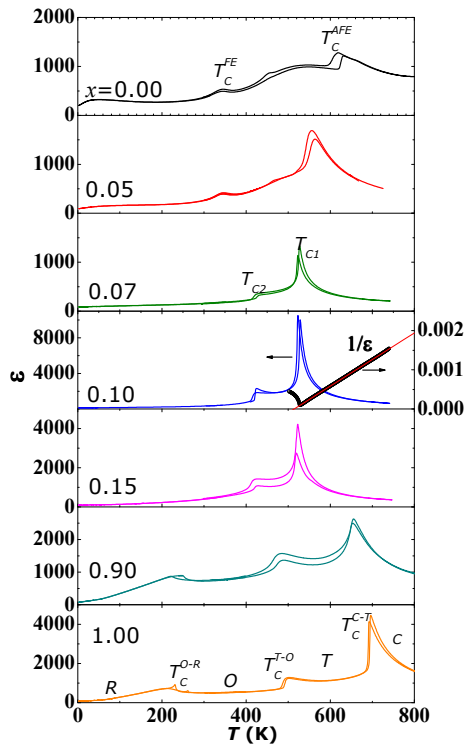

Fig. 20. Temperature dependence of dielectric constant for $Ag_{1-x}K_xNbO_3$ solid solution.

## 6. AgTaO₃

AgTaO₃ is another oxide of the two discovered silver perovskites (Francombe & Lewis, 1958). It is generally accepted that AgTaO₃ undergoes a series of phase transitions from rhombohedral phase ($T \leq 685$ K) to monoclinic phase (650 K$\leq T \leq 703$ K) and then to tetragonal phase (685 K$<T \leq 780$ K), and finally to cubic phase at $T_{\text{T-C}}$= 780 K upon heating (Francombe & Lewis, 1958; Kania, 1983; Paweczyk, 1987; Kugel et al.,1987; Hafid et al., 1992). However, due to the coexistence regions between rhombohedral and monoclinic, and monoclinic and tetragonal, the actual transition temperatures from rhombohedral to monoclinic $T_{\text{R-M}}$ as well as that from monoclinic to tetragonal $T_{\text{M-T}}$ still remain uncertain. Furthermore, the ground state and the origins that trigger these phase transitions still remain to be addressed (Soon et al., 2010; Kugel et al.,1987; Wołcyrz & Łukaszewski,1986).

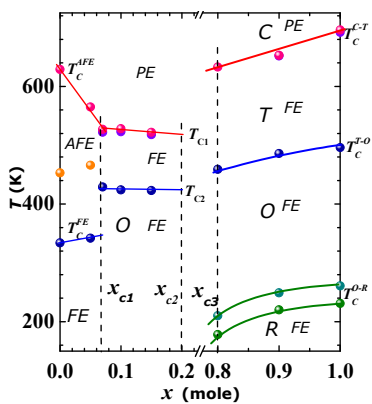

Fig. 21. Phase diagram proposed for the Ag$_{1-x}$K$_x$NbO$_3$ solid solution. Phase boundaries are indicated by the dashed lines. *C, T, O,* and *R* indicate the cubic, tetragonal, orthorhombic, and rhombohedral symmetries, respectively. PE, FE and AFE represent paraelectric, ferroelectric, and antiferroelectric phases, respectively.

### 6.1 Synthesis

Although single crystal of AgTaO₃ is available through a flux method (Łukaszewski et al., 1980; Kania, 1989), it is extremely difficult to prepare its dense ceramics sample (Francombe & Lewis, 1958;Kania,1983) for electrical measurements. Since decomposition of AgTaO₃ occurs at 1443±3 K in atmosphere (Valant et al., 2007b), sintering cannot be performed at higher temperatures to obtain dense ceramics. However, this long-standing synthesis difficulty now can be solved by a processing route involving the conventional solid-state reaction and sintering in environment with a high oxygen pressure at ~13 atm (Soon et al., 2010). In this synthesizing route, Ag₂O and Ta₂O₅, first underwent a grind mixing and was calcined at 1273 K for 6 hours. The calcined powder was then pressed into a pellet in 6 mm in diameter. Sintering was carried out by placing the powder compact into a sealed zirconia tube that was connected to a pressure control valve (Fig.22). Prior to the sintering, oxygen gas at ~6.25 atm was filled into the sealed zirconia tube after the evacuation. Upon heating, the pressure of sealed oxygen gas increased and reached ~13 atm when the powder compact was sintered at 1573K for 2 hours. This eventually led to formation of dense polycrystalline AgTaO₃.

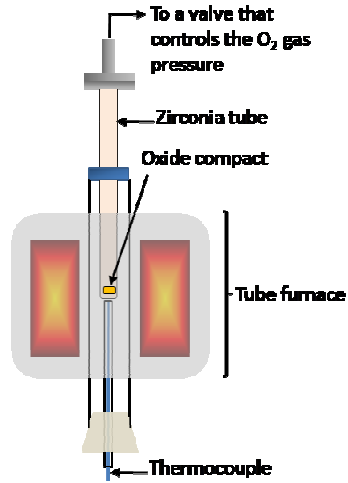

Fig. 22. Schematic diagram of the self-customized furnace employed for the sintering at high oxygen pressure.

## 6.2 Phase formation and dielectric behaviors

X-ray diffraction analyses (Fig.23) suggest the AgTaO₃ is rhombohedral with $R\bar{3}c$ symmetry at room temperature (Wołcyrz & Łukaszewski, 1986). Diffraction patterns obtained at 68.4 K remain unchanged, indicating that such nonpolar phase persists down to low-temperatures. This is also supported by the measurements on the dielectric constants (Fig.24) and heat capacity (Fig.25), in which no anomaly was probed in the low temperatures.

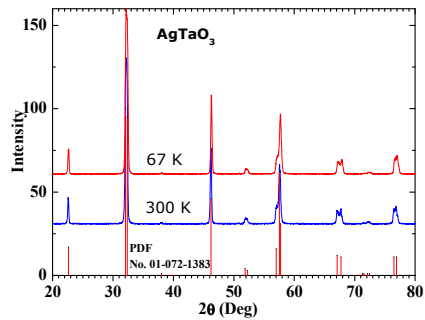

Fig. 23. XRD traces for AgTaO₃ obtained at 300 K and 68.4 K together with the standard pattern given by the powder diffraction file (PDF) No. 01-072-1383.

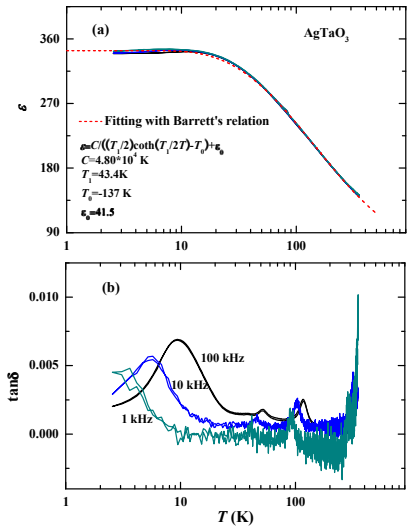

Fig. 24. Temperature dependence of (a) $\varepsilon$ together with the fitting to the Barrett's relation indicated by the dashed line and (b) $\tan\delta$ for AgTaO₃ ceramics samples.

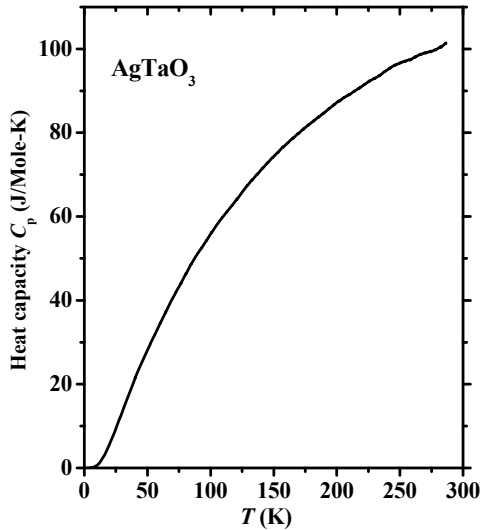

Fig. 25. Heat capacity for AgTaO₃.

Although several frequency-dependent peaks are seen in the dielectric loss (Fig.24(b)), it can be reasonably attributed to polarization relaxations due to defects (Soon et al., 2010). Interestingly, within the low-temperature region, the dielectric behavior follows the Barrett's relation (Barrett, 1952) that is characteristic for the quantum paraelectric system (Abel, 1971; Höchli & Boatner,1979; Itoh et al., 1999), suggesting that AgTaO₃ may be a

1   quantum paraelectric. On the other hand, two step-like dielectric anomalies corresponding
2   to the phase transitions from monoclinic to tetragonal and tetragonal to cubic were observed
3   at 694 and 780 K, respectively, upon heating the samples (Fig.26). This observation is in
4   agreement with the previous reports (Kania,1983;Kugel et al., 1987). Furthermore, the
5   temperature dependence of $1/\varepsilon$ for AgTaO₃ shows non-linear behavior, which is similar to
6   that of KTaO₃, obeying the modified form of the Curie-Weiss law $\varepsilon=\varepsilon_L+C/(T-T_0)$ (Rupprecht
7   & Bell, 1964).
8

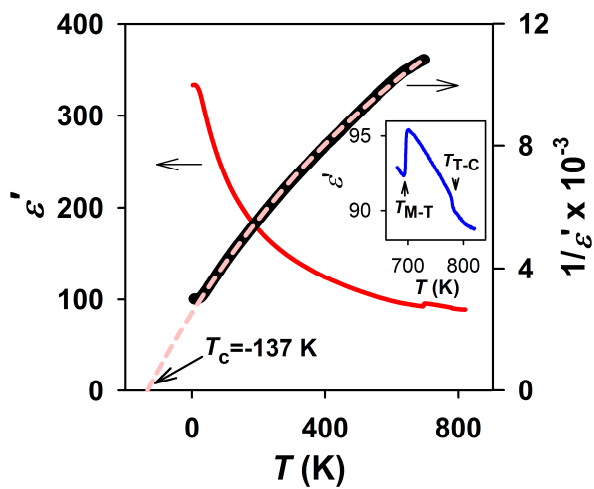

10  Fig. 26. Temperature dependences of $\varepsilon'$ and $1/\varepsilon'$ fitted to the modified Curie-Weiss law
11  $\varepsilon=\varepsilon_L+C/(T-T_0)$ (dashed line, $\varepsilon_L$= 38.2, C=4.7*10⁴ K, $T_0$= 137 K) for AgTaO₃ measured at 1
12  MHz upon heating. The inset further shows the step-like phase transitions from monoclinic
13  to tetragonal and then to cubic at $T_{M-T}$= 694 K and $T_{T-C}$ =780 K, respectively.

## 7. (Ag₁₋ₓLiₓ)TaO₃ solid solution

15  Similar to the case of AgNbO₃, ~12 mole% of Li can be incorporated into the Ag-site of
16  AgTaO₃ to form (Ag₁₋ₓLiₓ)TaO₃ (Soon et al., 2009). Although the transition temperature is
17  lower than room temperature, ferroelectricity can be induced in this solid solution due to
18  the strong off-centering nature of the small Li ions.

### 7.1 Synthesis

20  (Ag₁₋ₓLiₓ)TaO₃ was prepared by the conventional solid-state reaction with Ag₂O, Ta₂O₅ and
21  Li₂CO₃, where the powder mixture was calcined at 1273 K for 6 hours. The calcined powder
22  was then pressed into a pellet with 6 mm in diameter. Sintering was carried out by the same
23  high-pressure process used for pure AgTaO₃, which eventually led to formation of dense
24  ceramics samples of (Ag₁₋ₓLiₓ)TaO₃

### 7.2 Dielectric behaviours and confirmation of ferroelectric phase

26  Figures 27 & 28 plot the temperature dependence of dielectric constant $\varepsilon$ and loss $\tan\delta$ for
27  (Ag₁₋ₓLiₓ)TaO₃ with $x\leq$ 0.12 obtained at frequencies ranging from 1 kHz to 1 MHz,

respectively. It can be seen that a dielectric peak was gradually induced by $Li^+$ substitution in $AgTaO_3$. In contrast to the single peak of the dielectric constant, there are two to four peaks of the dielectric loss within the same temperature window. Since the additional loss peaks do not associate with a remarkable change in the dielectric constant, it is very likely due to the defect effects dependent with sample processing. It can be seen that a well-defined peaks has occurred in the ~MHz frequency regions for the substitution at extremely low level, for example, at $x$=0.02. This indicates the existence of local polarization in the doped crystal (Vugmeister & Glinchuk, 1990; Samara, 2003). This fact again suggests that $AgTaO_3$ actually under a critical state of the quantum paraelectric. Any slight modification will lead to the appearance of observable polarization in the system. For small substitution, the location of the dielectric anomaly depends on the observed frequency. Figure 29 gives an evaluation on this frequency dependence. In the figure, the temperature-axis is scaled with the peak position of 1 MHz, making it easy to see the change with composition. For $x$=0.008, the frequency dispersion is very strong, peak position of the dielectric constant shifts about 50% with respect to that of 1 MHz for $f$= 1 kHz. However, for $x$=0.035, such change is less than 2%, meaning that ferroelectric phase transition temperature $T_c$ is well defined in the sample. Thus, we can infer that a macroscopic ferroelectric phase is evolved below $T_c$ in this composition. This was also confirmed by the results shown in Fig.30, in which ferroelectric loop was obtained for $T < T_c$ for a sample with $x$=0.12 that has the same dielectric behaviors to that of $x$=0.035.

On the basis of the above results, a phase diagram is proposed in Fig.31, in which PE and FE represents the paraelectric and ferroelectric phases, respectively.  As mentioned above, since the peak of the dielectric constant is strongly dependent with frequency for $0 < x < 0.035$, the gray zone in the phase diagram may be attributed to dipole-glass phase(Vugmeister & Glinchuk, 1990;Pirc & Blinc, 1999;Samara, 2003) or a phase with nanosized ferroelectric domains (Fisch, 2003; Fu et al., 2009b), which remains to be addressed by further investigations.

## 8. Concluding remarks

Our recent works reveal that silver perovskites are of great interests from either the view-point of fundamental research or that of application research in the fields of ferroelectric or piezoelectric. Promising ferroelectric and piezoelectric properties have been demonstrated in some compounds such as $(Ag,Li)NbO_3$ and  $(Ag,K)NbO_3$ alloys, but further works are required to improve the material performance, to understand the basic physics of the ferroelectricity/piezoelectricity of the materials, and to seek novel promising compound among the discovered solutions or alloys with other ferroelectric systems. Moreover, integration techniques of thin films are also a direction for the future work when considering the practical applications.

## 9. Acknowledgment

Part of this work was supported by the Collaborative Research Project of Materials and Structures Laboratory of Tokyo Institute of Technology, and Grant-in-Aid for Scientific Research, MEXT, Japan.

1
2
3
4
5
6
7
8
9
10

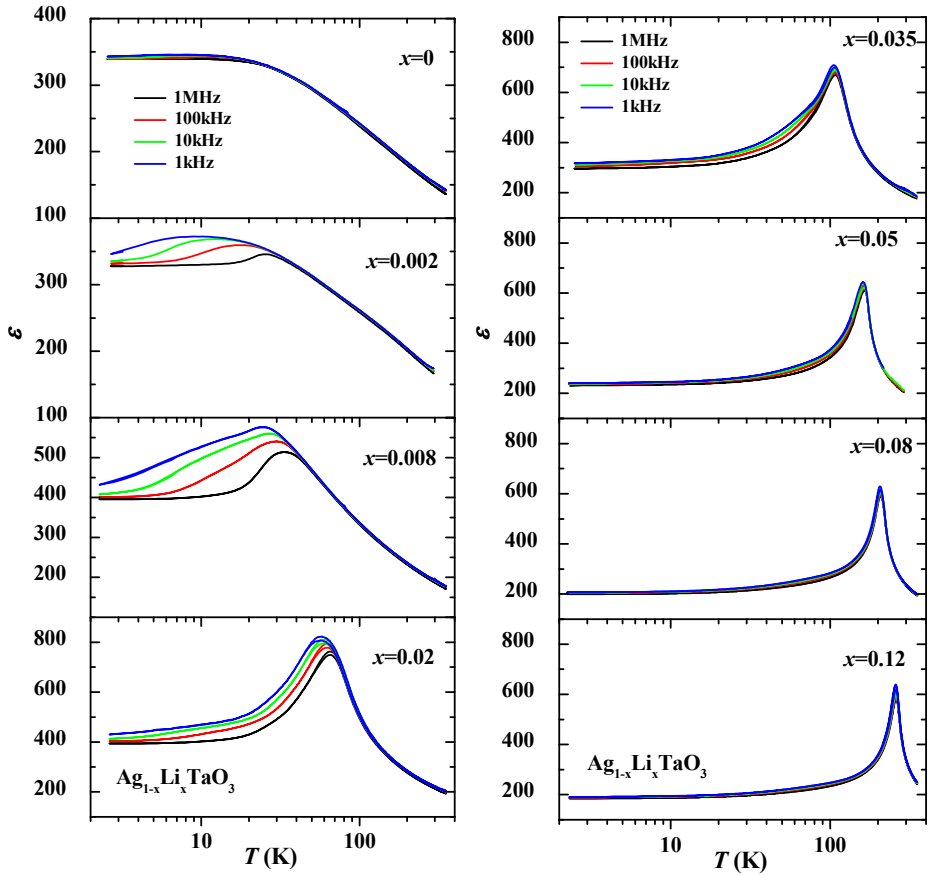

11
12
13
14
15
16
17
18 Fig. 27. Dielectric constant $\varepsilon'(T)$ for $Ag_{1-x}Li_xTaO_3$ with $0 \leq x \leq 0.12$.

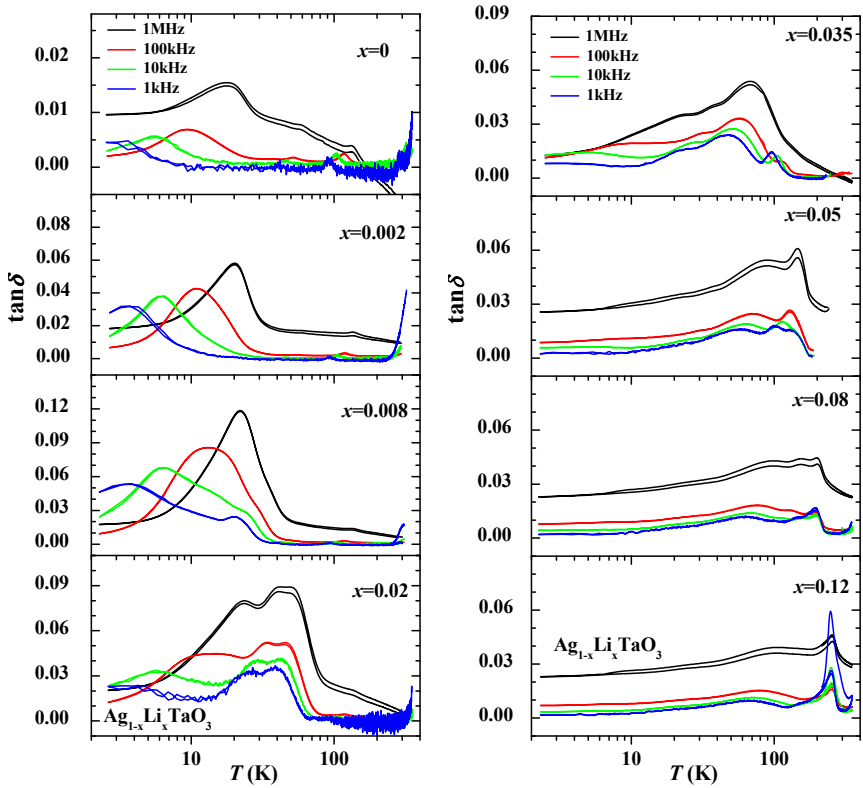

Fig. 28. Dielectric loss for $Ag_{1-x}Li_xTaO_3$ with $0 \leq x \leq 0.12$.

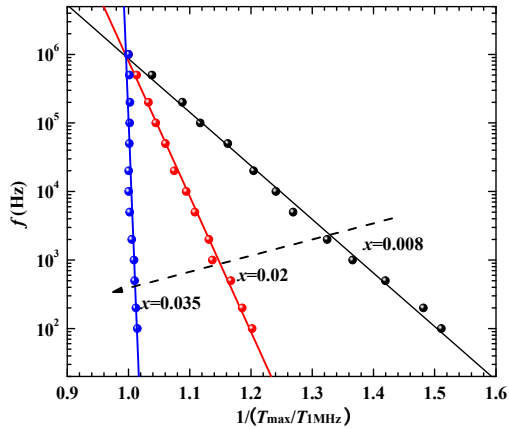

Fig. 29. Relationship between frequency and $T$max in $Ag_{1-x}Li_xTaO_3$. $T$max was normalized by the $T$max of 1 MHz.

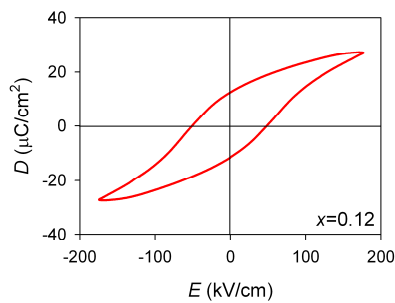

Fig. 30. Hysteretic $D$-$E$ loop for $Ag_{1-x}Li_xTaO_3$ with $x$=0.12 obtained at 0.1 Hz and 77 K, indicating the ferroelectric state at $T$< $T_c$ (=258 K).

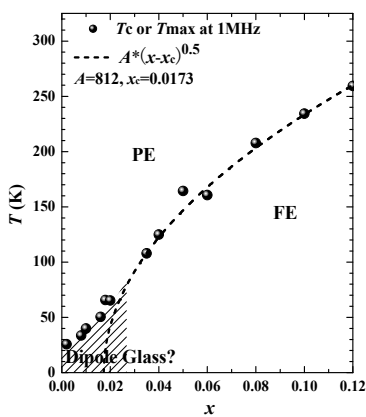

Fig. 31. Proposed phase diagram for $Ag_{1-x}Li_xTaO_3$. The dashed-dotted line shows the fitting to Morf-Schneider's relationship that is proposed for a quantum phase transition (Morf et al., 1977). PE and FE denote paraelectric and ferroelectric states, respectively. The gray zone may be a dipolar-glass state (Pirc & Blinc, 1999) or a state with nanosized ferroelectric domains (Fisch, 2003; Fu et al., 2009b).

## 10. References

Abel, W. R. (1971). Effect of pressure on the static dielectric constant of KTaO₃, *Phys. Rev. B* 4, pp. 2696-2701

Arioka, T. (2009 ) *Undergraduate thesis of Tokyo Institute of Technology*, Tokyo

Arioka, T.; Taniguchi, H.; Itoh, M.; Oka, K.; Wang, R. & Fu, D. (2010). Ferroelectricity in NaNbO₃: Revisited, *Ferroelectrics* 401, pp. 51-55

Baettig, P.; Ederer, C. & Spaldin, N. A. (2005). First principles study of the multiferroics BiFeO₃, Bi₂FeCrO₆, and BiCrO₃: Structure, polarization, and magnetic ordering temperature, *Phys. Rev. B* 72, pp. 214105

Barrett, J. H. (1952). Dielectric Constant in perovskite type crystals, *Phys. Rev.* 86, pp. 118-120

Bilc, D. I. & Singh, D. J. (2006). Frustration of tilts and A-site driven ferroelectricity in $KNbO_3$-$LiNbO_3$ alloys, *Phys. Rev. Lett.* 96, pp. 147602

Cohen, R. E. (1992). Origin of ferroelectricity in perovskite oxides, *Nature* 358, pp. 136–138

Eglitis, R. I.; Postnikov, A. V.& Borstel, G. (1997). Semiempirical Hartree-Fock calculations for pure and Li-doped $KTaO_3$, *Phys. Rev. B* 55, pp. 12976-12981

Egami, T.; Dmowski, W.; Akbas M. & Davies, P. K. (1998). Local structure and polarization in Pb containing ferroelectric oxides in *First-Principles Calculations for Ferroelectrics – 5th Williamsburg Workshop* ed Cohen R., pp 1–10, AIP, New York

Fabry, J.; Zikmund, Z.; Kania, A. & Petricek, V. (2000).  Silver niobium trioxide, $AgNbO_3$, *Acta Cryst.* C56, pp. 916-918

Fisch, R. (2003). Random-field models for relaxor ferroelectric behavior, *Phys. Rev. B* 67, pp. 094110

Fu, D.; Endo, M.; Taniguchi, H.; Taniyama, T. & Itoh, M. (2007). $AgNbO_3$: A lead-free material with large polarization and electromechanical response, *Appl. Phys. Lett.* 90, pp. 252907

Fu, D.; Endo, M.; Taniguchi, H.; Taniyama, T.; Koshihara, S. & Itoh, M. (2008) Piezoelectric properties of lithium modified silver niobate perovskite single crystals, *Appl. Phys. Lett.* 92, pp. 172905

Fu, D.; Itoh, M. & Koshihara, S. (2009a). Dielectric, ferroelectric, and piezoelectric behaviors of $AgNbO_3$–$KNbO_3$ solid solution, *J. Appl. Phys.* 106, pp. 104104

Fu, D.; Taniguchi, H.; Itoh, M.; Koshihara, S.; Yamamoto, N. & Mori, S. (2009b). Relaxor $Pb(Mg_{1/3}Nb_{2/3})O_3$: A ferroelectric with multiple inhomogeneities, *Phys. Rev. Lett.* 103, pp. 207601.

Fu, D.; Itoh, M. & Koshihara, S. (2010). Invariant lattice strain and polarization in $BaTiO_3$–$CaTiO_3$ ferroelectric alloys, *J. Phys.: Condens. Matter* 22, pp. 052204

Fu, D.; Endo, M.; Taniguchi, H.; Taniyama, T.; Itoh, M. & Koshihara, S. (2011). Ferroelectricity of Li-doped silver niobate $(Ag,Li)NbO_3$, *J. Phys.: Condens. Matter* 23, pp. 075901

Fortin, W. & Kugel, G. E.; Grigas, J. & Kania, A. (1996). Manifestation of Nb dynamics in Raman, microwave, and infrared spectra of the $AgTaO_3$-$AgNbO_3$ mixed system, *J. Appl. Phys.* 79, pp. 4273-4282

Francombe, M. H. & Lewis, B. (1958). Structural and electrical properties of silver niobate and silver tantalate, *Acta Cryst.* 11, pp. 175-178

Goldschmidt, V.M. (1926). *Shrifter Norske Videnskaps-Akad. Oslo, I: Mat.-Naturv. Kl.* No. 2, pp. 8

Grinberg, I. &Rappe, A. M. (2003). Ab initio study of silver niobate in *Fundamental Physics of Ferroelectrics*, edited by Davies, P. K. & Singh, D. J., pp 130–138,American Institute of Physics, New York

Grinberg, I. & Rappe, A. M. (2004). Silver solid solution piezoelectrics, *Appl. Phys. Lett.* 85, pp. 1760-1762

Hafid, M.; Kugel, G. E.; Kania, A.; Roleder, K. & Fontana, M. D. (1992). Study of the phase-transition sequence of mixed silver tantalate niobate $(AgTa_{1-x}Nb_xO_3)$ by inelastic light-scattering, *J. Phys. Conden. Matter.* 4, pp. 2333-2345

Höchli, U. T. & Boatner, L. A. (1979). Quantum ferroelectricity in $K_{1-x}Na_xTaO_3$ and $KTa_{1-y}Nb_yO_3$, *Phys. Rev B.* 20, pp. 266-275

Itoh, M.; Wang, R.; Inaguma, Y.; Yamaguchi, T.; Shan, Y-J. &Nakamura, T. (1999). Ferroelectricity induced by oxygen isotope exchange in strontium titanate perovskite, *Phys. Rev. Lett.* 82, pp. 3540-3543

Jaffe, B.; Cook, Jr., W. R. & Jaffe, H. (1971). *Piezoelectric Ceramics,* Academic, London

Kania, A. (1983). $AgNb_{1-x}Ta_xO_3$ solid solutions-dielectric properties and phase transitions, *Phase Transit.* 3, pp. 131-139

Kania, A.; Roleder, K. & Lukaszewski, M. (1984). The ferroelectric phase in $AgNbO_3$, Ferroelectrics 52, pp. 265-269

Kania, A.; Roleder, K.; Kugel, G. E. & Fontana, M. D. (1986). Raman scattering, central peak and phase transitions in $AgNbO_3$, *J. Phys. C: Solid State Phys.* 19, pp. 9-20

Kania, A. (1989). Flux growth of $AgTa_xNb_{1-x}O_3$ (ATN) solid solution single crystals, *J. Cryst. Growth* 96, pp. 703-704

Kania, A. (1998). An additional phase transition in silver niobate $AgNbO_3$, *Ferroelectrics* 205, pp. 19-28

Kania, A. & Kwapulinski, J. (1999). $Ag_{1-x}Na_xNbO_3$ (ANN) solid solutions: from disordered antiferroelectric $AgNbO_3$ to normal antiferroelectric $NaNbO_3$, *J. Phys.: Condens. Matter.*11, pp. 8933-8946

Kania, A. (2001). Dielectric properties of $Ag_{1-x}A_xNbO_3$ (A: K, Na and Li) and $AgNb_{1-x}Ta_xO_3$ solid solutions in the vicinity of diffuse phase transitions, *J. Phys. D Appl. Phys.*34, pp. 1447-1455

Kato, H.; Kobayashi, H. & Kudo, A. (2002). Role of $Ag^+$ in the band structures and photocatalytic properties of $AgMO_3$ (M: Ta and Nb) with the perovskite structure, *J. Phys. Chem. B* 106, pp. 12441-12447

Kruczek, M.; Talik, E. & Kania, A. (2006). Electronic structure of $AgNbO_3$ and $NaNbO_3$ studied by X-ray photoelectron spectroscopy, *Solid State Commun.* 137, pp. 469-473

Kugel, G. E.; Fontana, M. D.; Hafid, M.; Roleder, K.; Kania, A. & Pawelczyk, M. (1987). A Raman study of silver tantalate ($AgTaO_3$) and its structural phase transition sequence, *J. Phys. C: Solid State Phys.* 20, pp. 1217-1230

Kuroiwa, Y.; Aoyagi, S.; Sawada, A.; Harada, J.; Nishibori, E.; Takata, M. & Sakata, M. (2001). Evidence for Pb-O Covalency in Tetragonal $PbTiO_3$, *Phys. Rev. Lett.* 87, pp. 217601

Levin, I.; Krayzman, V.; Woicik, J. C.; Karapetrova, J.; Proffen, T.; Tucker, M. G. & Reaney, I. M. (2009). Structural changes underlying the diffuse dielectric response in $AgNbO_3$, *Phys. Rev.* B 79, pp. 104113

Lines, M. E. & Glass, A. M. (1977). *Principle and Application of Ferroelectrics and Related Materials,* Oxford, Clarendon

Łukaszewski, M.; Kania, A. & Ratuszna, A. (1980). Flux growth of single crystals of $AgNbO_3$ and $AgTaO_3$, *J. Cryst. Growth* 48, pp. 493-495

Łukaszewski, L.; Pawelczyk, M.; Handerek, J. & Kania, A. (1983). On the phase transitions in silver niobate $AgNbO_3$, *Phase Transit.* 3, pp. 247-258

Łukaszewski, L. (1983). Dielectric properties of $Ag_{1-x}K_xNbO_3$ solid solutions, *Ferroelectrics* 44, pp. 319-324

Mitchell, R. H. (2002). *Perovskites: modern and ancient,* Almaz press, Ontario

Morf, R.; Schneider, T. & Stoll, E. (1977). Nonuniversal critical behavior and its suppression by quantum fluctuations, *Phys. Rev. B* 16, pp. 462-469

Müller, K. A. & Burkard, H. (1979). $SrTiO_3$: An intrinsic quantum paraelectric below 4 K, *Phys. Rev. B* 19, pp. 3593-3602

Nalbandyan, V. B.; Medviediev, B. S. & Beliayev, I. N. (1980). Study of the silver niobate-lithium niobate systems, *Izv.Akad. Nauk SSSR, Neorg. Mater.* 16, pp. 1819-1823

Paweczyk, M. (1987). Phase transitions in $AgTa_xNb_{1-x}O_3$ solid solutions, *Phase Transit.* 8, pp. 273-292

Petzelt, J.; Kamba, S.; Buixaderas, E.; Bovtun, V.; Zikmund, Z.; Kania, A.; Koukal, V.; Pokorny, J.; Polivka, J.; Pashkov, V.; Komandin, G. & Volkov, A. (1999). Infrared and microwave dielectric response of the disordered antiferroelectric $Ag(Ta,Nb)O_3$ system, *Ferroelectrics* 223, pp. 235-246

Pirc, R. & Blinc, R. (1999). Spherical random-bond–random-field model of relaxor ferroelectrics, *Phys. Rev. B* 60, pp. 13470-13478

Pisarski, M. & Dmytrow, D. (1987). Phase transitions in ceramic $AgNbO_3$ investigated at high hydrostatic pressure, *Ferroelectrics* 74, pp. 87-93

Ratuszna, A.; Pawluk, J. &Kania, A. (2003). Temperature evolution of the crystal structure of $AgNbO_3$, *Phase Transit.* 76, pp. 611-620

Reisman, A. & Holtzberg, F. (1958). Heterogeneous equilibria in the systems $Li_2O$–, $Ag_2O$–$Nb_2O_5$ and oxide-models. *J. Am. Chem. Soc.* 80, pp. 6503-6507

Rodel, J. Jo, W.; Seifert, K. T. P.; Anton, E.; Granzow, T. & Damjanovic, D. (2009). Perspective on the development of lead-free piezoceramics, *J. Am. Ceram. Soc.* 92, pp. 1153 -1177

Rupprecht, G. & Bell, R. O. (1964). Dielectric constant in paraelectric perovskites, *Phys. Rev.* 135, pp. A748-A752

Saito, Y.; Takao, H.; Tani, T.; Nonoyama, T.; Takatori, K.; Homma, T.; Nagaya, T. & Nakamura, M. (2004). Lead-free piezoceramics, *Nature* 432, pp. 84-87

Sakabe, Y.; Takeda, T.; Ogiso, Y. & Wada, N. (2001). Ferroelectric Properties of (Ag, Li)$(Nb,Ta)O_3$ Ceramics, *Jpn. J. Appl. Phys. Part* 1, 42 pp. 5675-5678

Samara, G. A. (2003).The relaxational properties of compositionally disordered $ABO_3$ perovskite. *J. Phys.: Condens. Matter* 15, pp. R367-R411

Sano, R., Morikawa, D.; Tsuda, K.; Fu, D. & Itoh, M. (2010). Space group determination of $AgNbO_3$ at room temperature by CBED method, *Meeting abstracts of the physical society of Japan* 65 (issue 2, part 4), pp. 916

Sciau, Ph; Kania, A.; Dkhil, B.; Suard, E. & Ratuszna, A. (2004). Structural investigation of $AgNbO_3$ phases using x-ray and neutron diffraction, *J. Phys.: Condens. Matter* 16, pp. 2795-2810

Scott, J. F. (2000). *Ferroelectric Memories*, Springer , Berlin

Shannon, R. D. (1976). Revised effective ionic radii and systematic studies of interatomic distances in halides and chalcogenides, *Acta Crystallogr., Sect. A: Cryst. Phys., Diffr., Theor. Gen.Crystallogr.* 32, pp. 751-767

Shigemi, A.; Wada, T. (2008). Crystallographic phase stabilities and electronic structures in $AgNbO_3$ by first-principles calculation, *Molecular Simulation* 34, pp. 1105-1114

Shiozaki, Y.; Nakamura, E. & Mitsui, T. (2001). *Ferroelectrics and Related Substances*, Vol. 36, Pt. A1, Springer, Berlin

Soon, H. P.; Taniguchi, H. &Itoh, M. (2010). Dielectric and soft-mode behaviors of $AgTaO_3$, *Phys. Rev. B* 81, pp. 104105

Soon, H. P.; Taniguchi, H. & Itoh, M. (2009). Ferroelectricity triggered in the quantum paraelectric $AgTaO_3$ by Li-substitution, *Appl. Phys. Lett.* 95, pp. 242904

Takeda, T.; Takahashi, Y.; Wada, N. & Sakabe, Y.  (2003). Effects of substitution of Na and K ions for Ag ion in (Ag,Li)NbO$_3$ Ceramics,  *Jpn. J. Appl. Phys.* Part 1, 42, pp. 6023-6026

Uchino, K. (1997). *Piezoelectric Actuators and Ultrasonic Motors*, Kluwer Academic, Boston

Valant, M.; Axelsson, A.; Alford, N. (2007a). Review of Ag(Nb,Ta)O$_3$ as a functional material, *J. Euro. Ceram. Soc.* 27, pp. 2549-2560

Valant, M.; Axelsson, A.; Zou, B. & Alford, N. (2007b). Oxygen transport during formation and decomposition of AgNbO$_3$ and AgTaO$_3$, *J. Mater. Res.* 22, pp. 1650-1655

Verwerft, M.; Dyck, D. V.; Brabers, V. A. M.; Landuyt, J. V. & Amelinckx S. (1989). Electron microscopic study of the phase transformations in AgNbO$_3$, *Phys. Stat. Sol. (a)* 112, pp. 451-466

Volkov, A. A.; Gorshunov, B. P.; Komandin, G.; Fortin, W.; Kugel, G. E.; Kania, A. & Grigas, J. (1995). High-frequency dielectric spectra of AgTaO$_3$-AgNbO$_3$ mixed ceramics, *J. Phys.: Condens. Matter* 7, pp. 785-793

Vugmeister, B. E. & Glinchuk, M. D. (1990). Dipole glass and ferroelectricity in random-site electric dipole systems, *Rev. Mod. Phys.* 62, pp. 993-1026

Wada, S., Park, S.-E.; Cross, L. E. & Shrout, T. R. (1998). Domain configuration and ferroelectric related properties of relaxor based single crystals, *J. Korean Phys. Soc.* 32, pp. S1290-S1293

Weirauch, D. F. &Tennery, V. (1967). Electrical, X-Ray, and thermal expansion studies in the system KNbO$_3$-AgNbO$_3$, *J. Am. Ceram. Soc.* 50, pp. 671-673

Wołcyrz, M. & Łukaszewski, M. (1986). The crystal structure of the room-temperature phase of AgTaO$_3$. *Zeitschrift für Kristallographie* 177, pp. 53-58

Yashima, M.; Matsuyama, S.; Sano, R.; Itoh, M.; Tsuda, K. & Fu, D. (2011). Structure of ferroelectric silver niobate AgNbO$_3$, *Chem. Mater.* 23, pp. 1643-1645